\documentclass[11pt,a4paper]{article}
\pdfoutput=1
\usepackage{jinstpub}
\usepackage{placeins}
\usepackage{amsmath}
\usepackage{caption}
\usepackage{subcaption}

\graphicspath{ {images/} }

\title{Commissioning of the new ALICE Inner Tracking System}
\collaboration[c]{on behalf of the ALICE ITS Collaboration}
\author[a,b]{J. P. Iddon}
\affiliation[a]{Department of Physics, University of Liverpool, Liverpool, UK}
\affiliation[b]{CERN, 1211 Geneva 23, Switzerland}
\emailAdd{james.philip.iddon@cern.ch}
\abstract{The upgrade of the Inner Tracking System (ITS) of ALICE (A Large Ion Collider Experiment) will extend measurements of heavy-flavour hadrons and low-mass dileptons to a lower transverse momentum than currently achieved and increase the readout capabilities to incorporate the full interaction. Furthermore, the tracking efficiency will be improved at low transverse momentum. To achieve this, the new ALICE ITS is comprised of seven layers of a custom Monolithic Active Pixel Sensor design known as ALPIDE, with a spatial resolution of $5\,\mathrm{\mu m}$. The use of the ALPIDE-based detector design will reduce the material budget to $0.35\%$\,$\mathrm{X_0}$ per layer for the innermost three layers, and to $1.0\%$\,$\mathrm{X_0}$ per layer for the outermost four layers, compared to $1.14\%$\,$\mathrm{X_0}$ per layer in the previous ITS. The construction effort in numerous sites around the world has resulted in a fully assembled and connected detector, which is currently undergoing on surface commissioning before its installation in the ALICE cavern. This contribution discusses the design and the current status of the commissioning of the new ITS detector, including the methods used to characterise the detector and the results obtained so far.}

\begin{document}
\maketitle
\flushbottom

\section{Introduction}
ALICE is the dedicated heavy ion experiment at CERN \cite{z}. The primary physics goal of ALICE is to study the state of matter at the highest energy densities reached in heavy-nucleus collisions. Under these conditions, colour confinement no longer occurs within the spatial confines of hadrons, meaning quarks and gluons are free to roam within the medium, known as the Quark Gluon Plasma (QGP). \par

The innermost sub-detector of ALICE is the Inner Tracking System (ITS). In Runs 1 and 2 of the Large Hadron Collider (LHC), the ITS consisted of two layers of Silicon Pixel Detectors (SPD), two layers of Silicon Strip Detectors (SSD) and two layers of Silicon Drift Detectors (SDD). For Run 3, an entirely new ITS has been constructed. The main goals of this upgrade are to improve the physics reach for low-mass dielectrons and rare probes at low transverse momentum ($p_{\mathrm{T}}$) by increasing vertex resolution, tracking efficiency and readout rate \cite{a}. An integrated luminosity of $13\,\mathrm{nb^{-1}}$ will be delivered during Run 3 for Pb-Pb collisions, which is a factor 100 larger compared to Runs 1 and 2 for minimum bias events.\par

\begin{figure}[htpb!]
  \centering
  \includegraphics[width=0.5\textwidth]{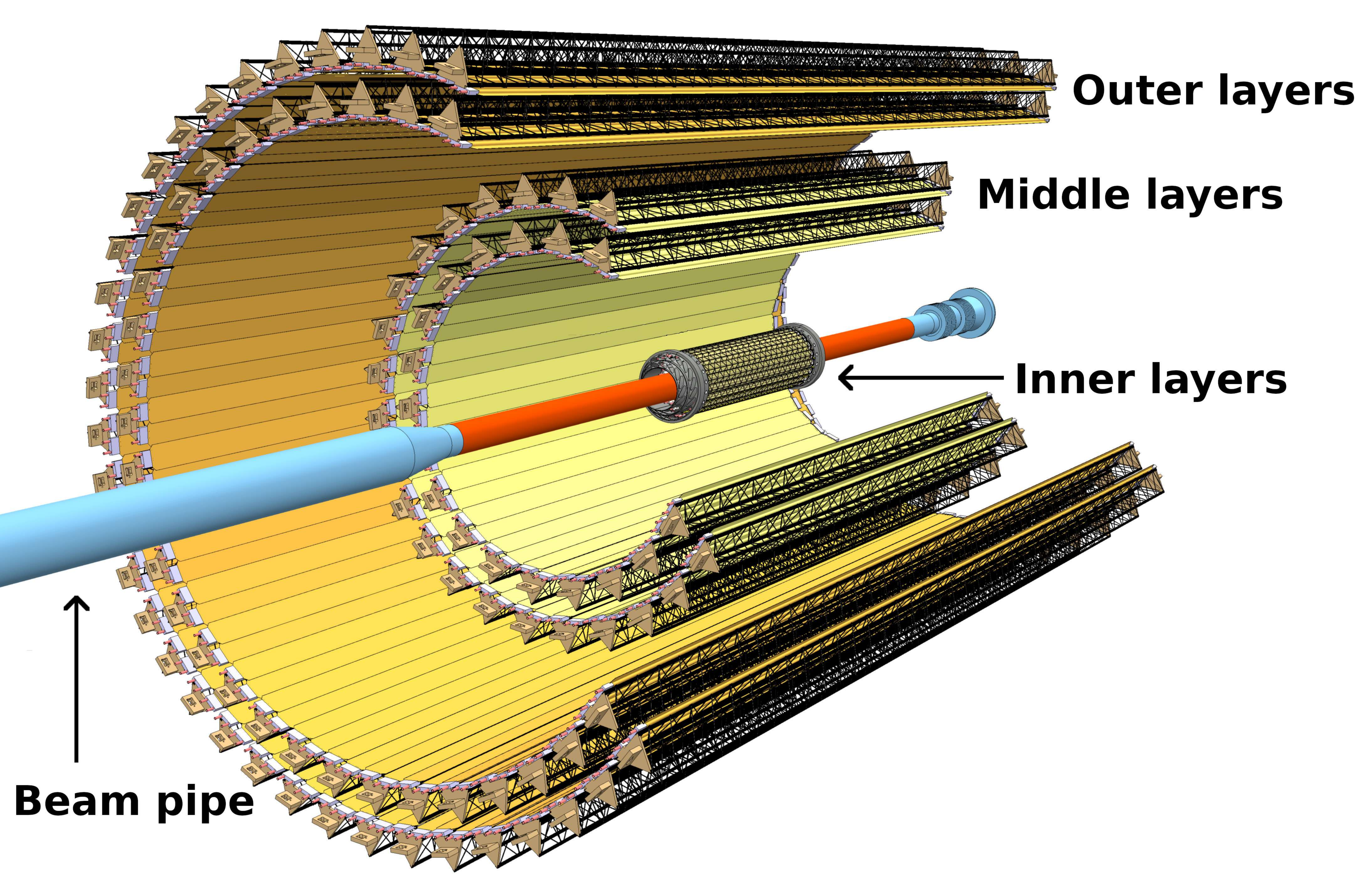}
  \caption{Layout of the new ITS. Taken from \cite{a}.}
  \label{fig:newITS}
\end{figure}

The new ITS, a schematic of which is shown in Fig.\,\ref{fig:newITS}, consists of seven layers of a CMOS Monolithic Active Pixel Sensor (MAPS) design known as ALice PIxel DEtector (ALPIDE) \cite{y} which is discussed in Section \ref{subsection:alpide}. The innermost three layers of the ITS are known as the Inner Barrel (IB), whilst the outermost four layers are known as the Outer Barrel (OB). The OB is further segmented into the Middle Layers (ML) and Outer Layers (OL) which are the innermost two layers of the OB and the outermost two layers of the OB respectively. \par

The readout rate of the upgraded ITS will be improved to $100\,\mathrm{kHz}$ which is twice the Pb-Pb interaction rate. In addition, the radius of the first layer of the ITS will be reduced from $39\,\mathrm{mm}$ to $23\,\mathrm{mm}$ and the pixel size reduced from $425\,\mathrm{\mu m} \times 50\,\mathrm{\mu m}$ in the SPD, to $\textit{\textbf{O}}(30\,\mathrm{\mu m})$ $\times$ $\textit{\textbf{O}}(30\,\mathrm{\mu m})$. The pseudorapidity region of |$\mathrm{\eta}$| $< 1.22$ for the 90\% most luminous area will be covered by the tracker, which has an active region of roughly $10\,\mathrm{m^2}$, segmented in 12.5~billion pixels.  

The upgrade will result in an improved impact parameter resolution, which will be reduced by a factor six in the direction along the beam axis (from $240\,\mathrm{\mu m}$ to $40\,\mathrm{\mu m}$) and by a factor three in the transverse plane (from $120\,\mathrm{\mu m}$ to $40\,\mathrm{\mu m}$) at a transverse momentum of $500\,\mathrm{MeV}$/$c$. An example of the anticipated performance of the upgraded ITS detector, with an integrated luminosity of $10\,\mathrm{nb^{-1}}$, is the measurement of the nuclear modification factor and anisotropic flow down to $p_{\mathrm{T}}$ of $2\,\mathrm{GeV}$/$c$ and $3\,\mathrm{MeV}$/$c$ respectively for the $\mathrm{\Lambda_c}$ baryon \cite{a}.\par

\FloatBarrier

\section{Detector design}
\subsection{The ALPIDE chip} \label{subsection:alpide}
\begin{figure}[htbp!]
  \centering
  \includegraphics[width=0.5\textwidth]{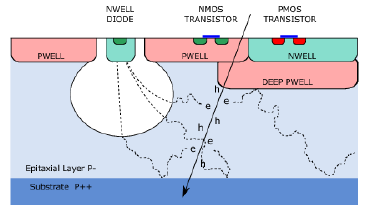}
  \caption{Cross section schematic of the ALPIDE chip. Taken from \cite{a}.}
  \label{fig:alpide}
\end{figure}

The ALPIDE chip, shown in Fig.\,\ref{fig:alpide}, is manufactured by TowerJazz with their $180\,\mathrm{nm}$ CMOS imaging process \cite{b}. The ALPIDE chip is $15\,\mathrm{mm} \times 30\,\mathrm{mm}$ and consists of $512 \times 1024$ pixels. Each pixel measures $27\,\mathrm{\mu m}$ $\times$ $29\,\mathrm{\mu m}$ and can be masked if necessary. The chips are thinned down to $50\,\mathrm{\mu m}$ and $100\,\mathrm{\mu m}$ for the IB and OB respectively. \par

A key feature of the design is the deep p-well shielding of the n-well, allowing the use of PMOS transistors inside the pixel matrix and therefore full CMOS logic. This makes it possible to have an amplifier, signal-shaper, discriminator and multiple event buffers in-pixel. The front-end is continuously active and has a power consumption of $40\,\mathrm{mW/cm^{2}}$. The pixels are arranged in double columns and read out by a priority encoder. Only the addresses of hit pixels are sent to the chip periphery.

A $2\,\mathrm{\mu m}$ diameter, low capacitance n-well diode is used together with an epitaxial layer resistance of $1\,\mathrm{k\Omega\!\cdot\!cm}$ and a reverse bias voltage of -$3\,\mathrm{V}$. This contributes to a radiation tolerance of $270\,\mathrm{krad}$~TID and $1.7\times10^{12}\,\mathrm{1MeV}$/$\mathrm{n_{eq}}$ NIEL which is the expected dose after 10 years of operation of the ITS. The n-well diode is roughly 300 times smaller than the pixel size, which combined with the reverse bias and hence reduced capacitance of the n-well diode, increases the signal to noise ratio. This is important for good detection efficiency at low power consumption.\par

The peaking time of the output of the front end is approximately $2\,\mathrm{\mu s}$, and the discriminated pulse has a duration of $5-10\,\mathrm{\mu s}$. For a more detailed insight into the ALPIDE chip, see \cite{c}.

\FloatBarrier

\subsection{Modules and staves}
\begin{figure}[htpb!]
  \centering
  \includegraphics[width=\textwidth]{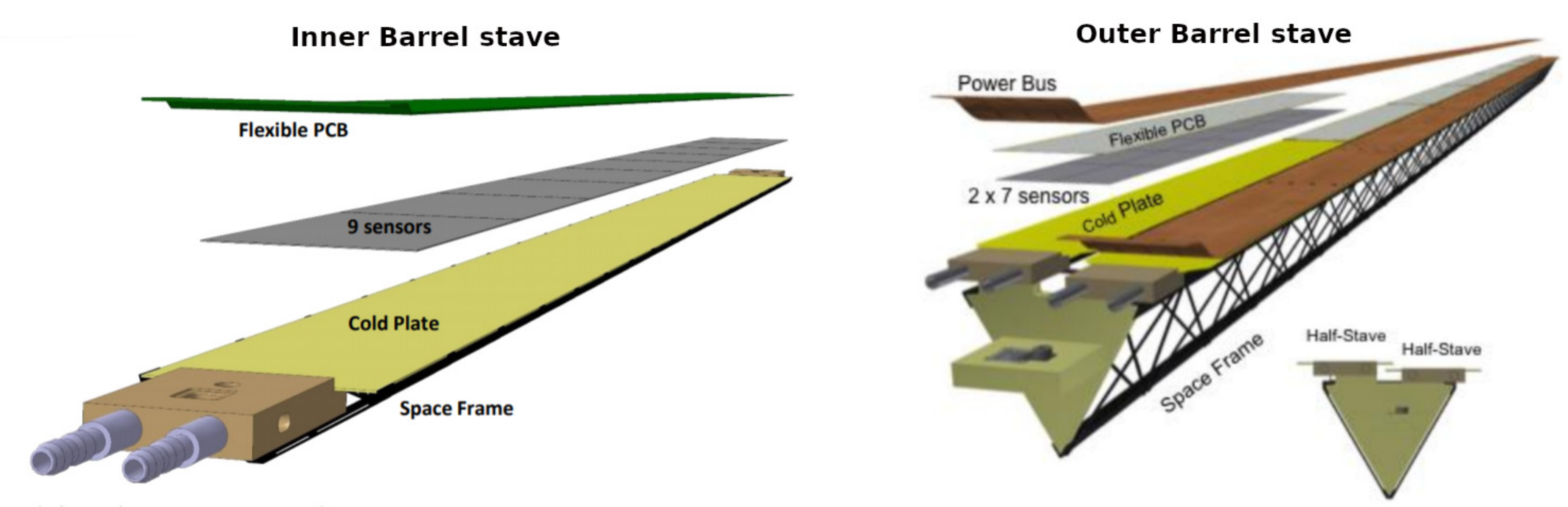}
  \caption{Schematic diagrams of an IB stave (left) and OL stave (right).}
  \label{fig:staves}
\end{figure}

A diagram of IB and OL staves can be seen in Fig.\,\ref{fig:staves}. Each IB stave consists of one IB module as well as a carbon fiber cold plate and a space frame. The cold plate has a continuous pipe running along the base, through which 20\,\textdegree{}C water is circulated. The full system is under atmospheric pressure and the outlet pressure is lower than the inlet pressure (a `leakless' system). Each IB module consists of 9 ALPIDE chips bonded structurally via glue and electrically via aluminium wirebonds to a Printed Circuit Board (PCB). The IB layers have an average radial position of $23\,\mathrm{mm}$, $31\,\mathrm{mm}$ and $39\,\mathrm{mm}$, and a total number of staves per layer of 12, 16 and 20 respectively. The weight of an IB stave is about $25\,\mathrm{g}$. \par

Each OL stave consists of two half staves that overlap. Each half stave consists of 7 OB modules, a power bus, as well as a carbon fiber cold plate and a space frame. ML staves have the same design as the OL staves but have 4 OB modules instead of 7. Each OB module consists of 14 ALPIDE chips arranged in two parallel rows of 7, bonded to a PCB. Power is distributed to each module via the power bus. The OB layers have an average radial position of $194\,\mathrm{mm}$, $247\,\mathrm{mm}$, $353\,\mathrm{mm}$ and $405\,\mathrm{mm}$, and a total number of staves per layer of 24, 30, 42 and 48 respectively. The total number of chips per stave is 112 for the ML and 196 for the OL. The weights of ML and OL staves is about $200\,\mathrm{g}$ and $400\,\mathrm{g}$ respectively. \par

Each chip in the IB is read out in parallel via its own high speed data link. In the OB, each module is split into two strips of 7 chips, which are segmented into 1 master and 6 slaves. Each master in the OB is read out in parallel via its own high speed link. The bandwidths of the IB and OB are $1.2\,\mathrm{Gbit/s}$ and $400\,\mathrm{Mbit/s}$ per link respectively. \par


\FloatBarrier

\begin{figure}[htpb!]
  \begin{subfigure}[b]{0.5\textwidth}
    \centering
    \includegraphics[width=\textwidth]{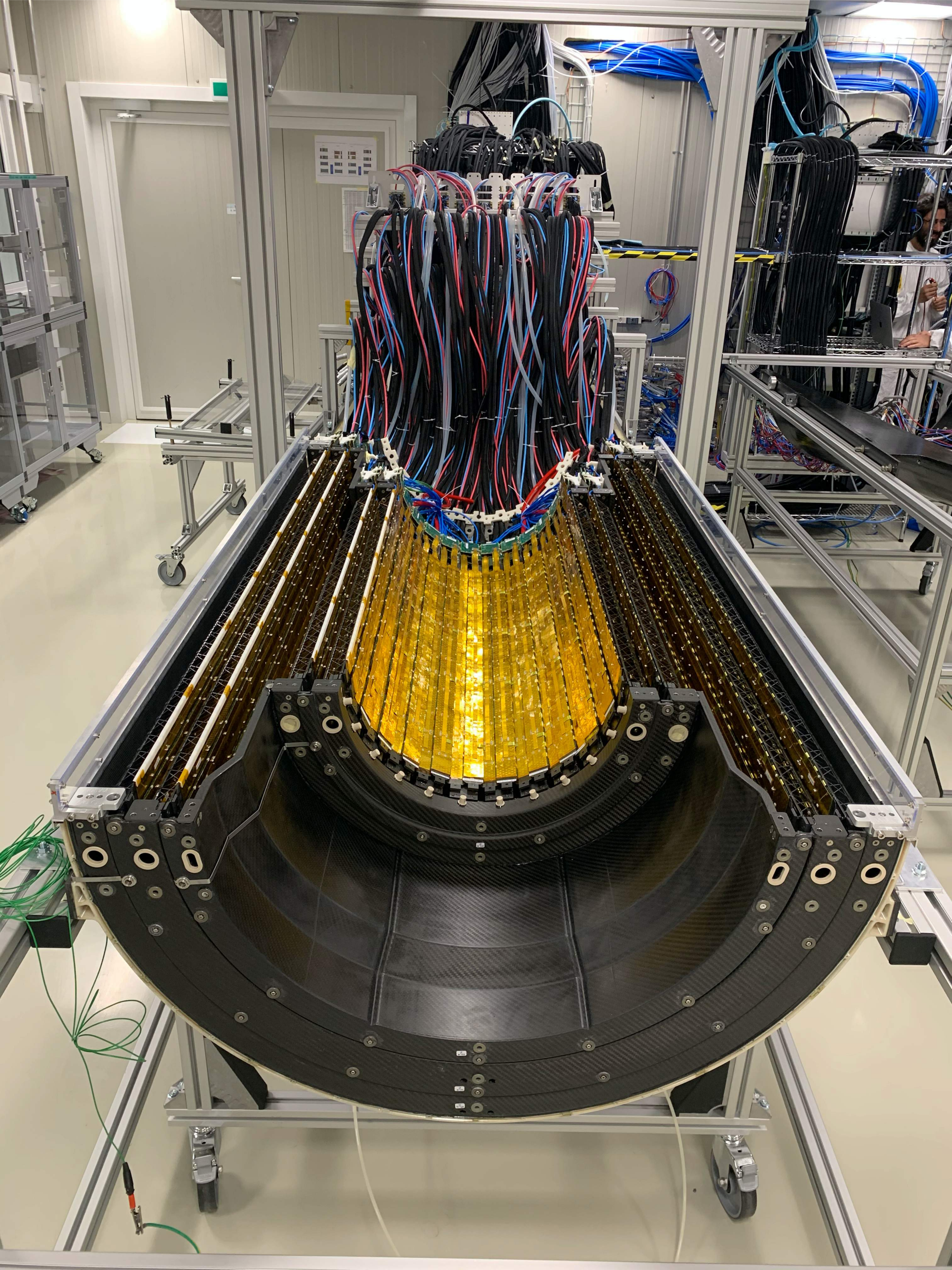}
    \caption{OB half barrel.}
    \label{fig:OBbottom}
  \end{subfigure}
  \begin{subfigure}[b]{0.5\textwidth}
    \centering
    \includegraphics[width=\textwidth]{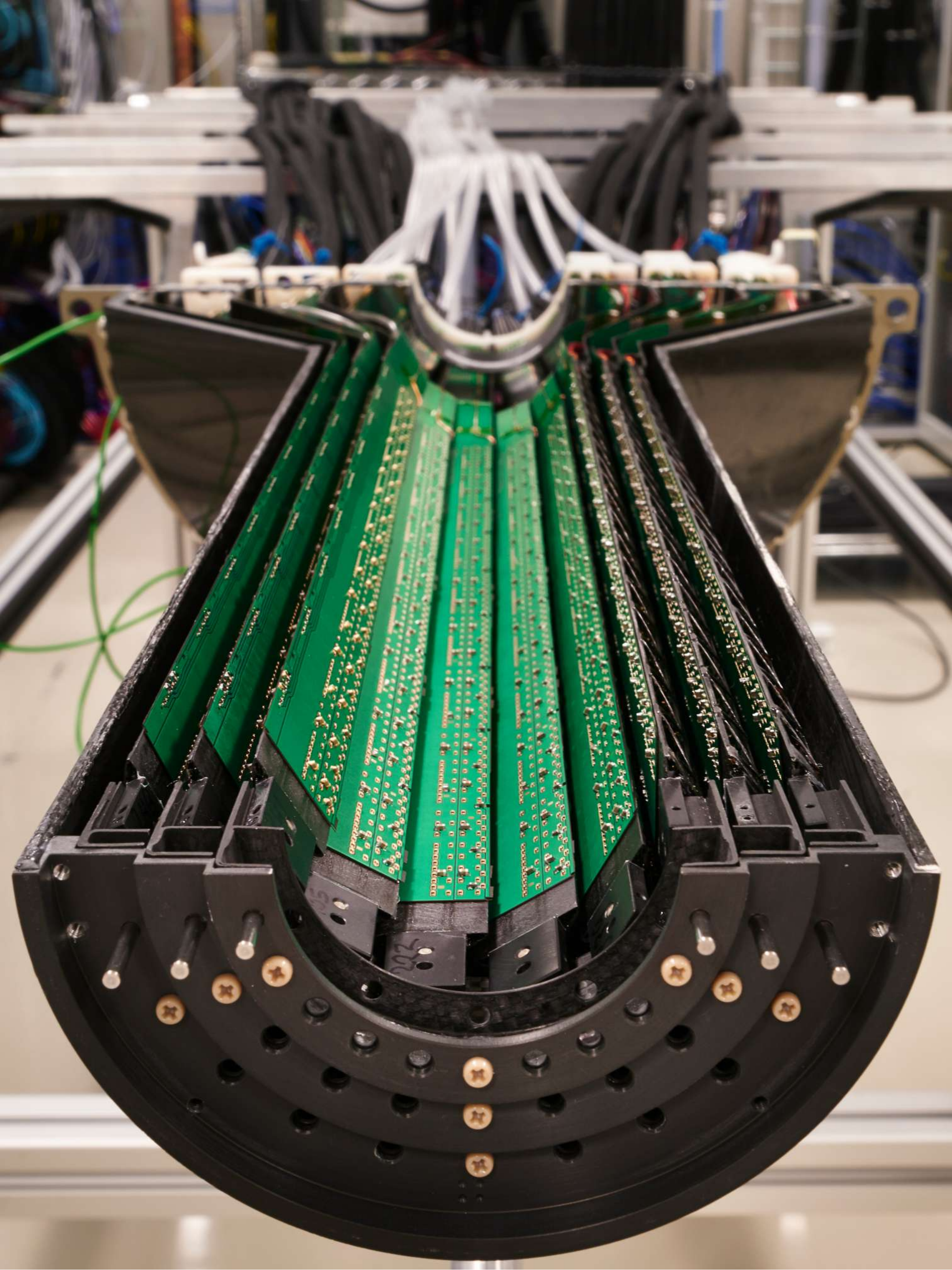}
    \caption{IB half barrel.}
    \label{fig:IBbarrel}
  \end{subfigure}
  \caption{One fully assembled ITS half barrel.}
\end{figure}

The construction of the IB, ML and OL including spares finished in July 2019, October 2019 and December 2019 respectively. The overall yield for stave construction was 73\% for the IB and 94\% for both the ML and OL. The assembly and connection to the readout system (discussed in Section \ref{section:readout}) and cooling of the IB and OB were completed in mid and late 2019 respectively. Figures \ref{fig:OBbottom} and \ref{fig:IBbarrel} show the fully assembled half barrels for the OB and IB respectively.

\FloatBarrier

\subsection{Readout System}
\label{section:readout}
\begin{figure}[htpb!]
  \centering
  \includegraphics[width=0.8\textwidth]{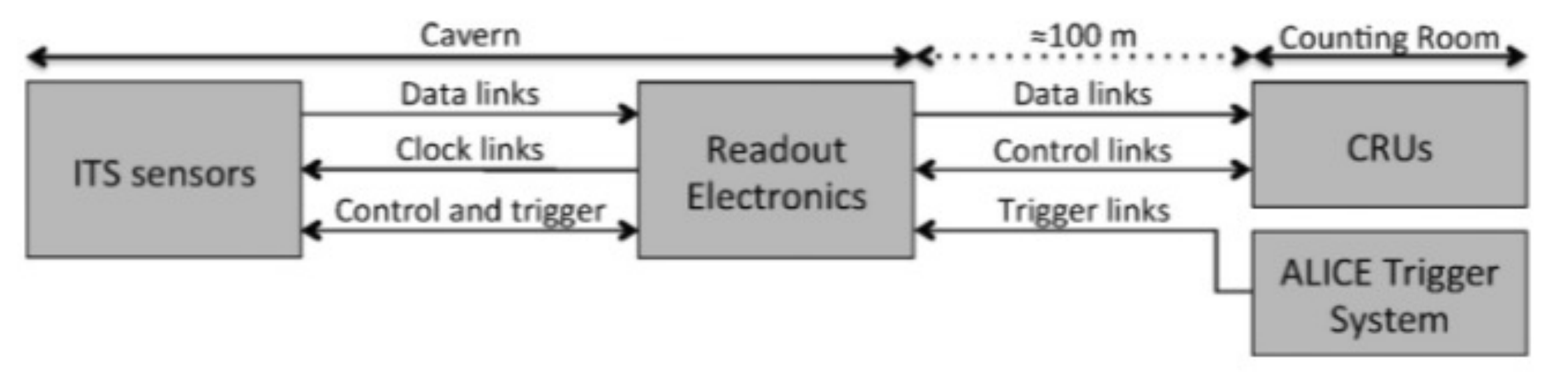}
  \caption{Architecture of the readout system. From \cite{e}.}
  \label{fig:ru}
\end{figure}

Each chip in the IB or each master chip in the OB receives clock and control signals from, and sends hit pixel data to, a Readout Unit (RU) via an $8\,\mathrm{m}$ long Samtec twinax copper differential link. Every stave is connected to a dedicated RU, a total of 192 units \cite{x}. The RUs are located in a lower radiation environment than the detector, resulting in less than $10\,\mathrm{krad}$ TID and $10^{11}\,1\,\mathrm{MeVn_{eq}/cm^2}$~NIEL. The RUs are connected via CERN GBT \cite{d} link to a Central Trigger Processor (CTP), from which they receive triggers, as well as a Central Readout Unit (CRU), to which data from the RUs are shipped. The CRUs are housed within the Front-end Level Processor computers which are in a background level radiation environment, the Counting Room. The detector is configured via the same datapath, which is shown in Fig.\,\ref{fig:ru}.

\FloatBarrier

\section{Outer Barrel commissioning}
Each stave in the ITS was tested with the full readout system after installation in the ITS. In the following, the focus lies on the commissioning of the OB. The basic test of each stave consisted of a verification of the power consumption, the control communication and the high speed links. After this, threshold tuning was conducted (see Section \ref{subsection:tune}). Finally, a fake-hit rate measurement was carried out (see Section \ref{subsection:fhr}).

\subsection{Outer Barrel threshold tuning}
\label{subsection:tune}
The threshold value of each chip is defined as the charge for which a pixel fires with a probability of 50\%. In ALPIDE, every pixel has an injection capacitor allowing the stimulation of the input of the front-end circuitry.\par

The threshold value of each chip can be adjusted chip-wide by augmenting two on-chip digital to analogue converters within the front end circuitry, VCASN and ITHR. The threshold is increased by decreasing VCASN or increasing ITHR \cite{a}. VCASN acts exponentially on the baseline, while ITHR acts in a linear fashion on the pulse height. For this reason, VCASN is used for the main tune and ITHR is used for a finer tune. \par

\begin{figure}[htbp!]
  \centering
  \includegraphics[width=\textwidth]{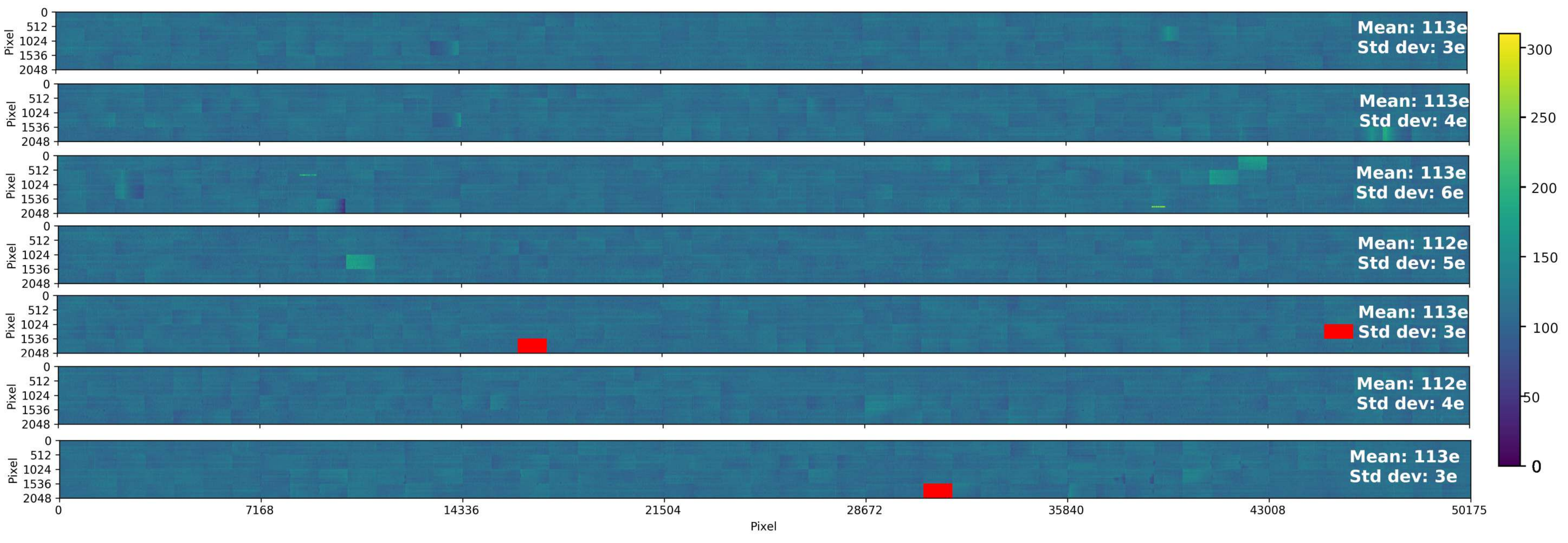}
  \caption{Tuned threshold scan of a selection of OL staves. Red chips are excluded from data taking. Threshold value units are number of electrons.}
  \label{fig:tune}
\end{figure}

For the threshold tuning, the concerned parameter is swept while a fixed charge corresponding to the target threshold is injected. The data are then analysed to see which parameter gives a threshold value closest to the injected charge. After this, a threshold scan is run with the optimum parameter. The threshold scan simply injects charge of increasing amplitude into the pixels, whilst keeping the chip configuration constant. Figure \ref{fig:tune} shows the threshold values of 7 randomly selected OL staves after tuning VCASN. The average threshold value over the staves shown after tuning is $112\,\mathrm{e^-}$, with a standard deviation of $4\,\mathrm{e^{-}}$. The standard deviation of threshold values from chip to chip within each stave varies from $3\,\mathrm{e^-}$ to $6\,\mathrm{e^-}$. To improve the threshold uniformity, scanning over ITHR can be performed. Likely this will further reduce the range in threshold values from chip to chip. \par

\begin{figure}[htbp!]
  \centering
  \includegraphics[width=0.5\textwidth]{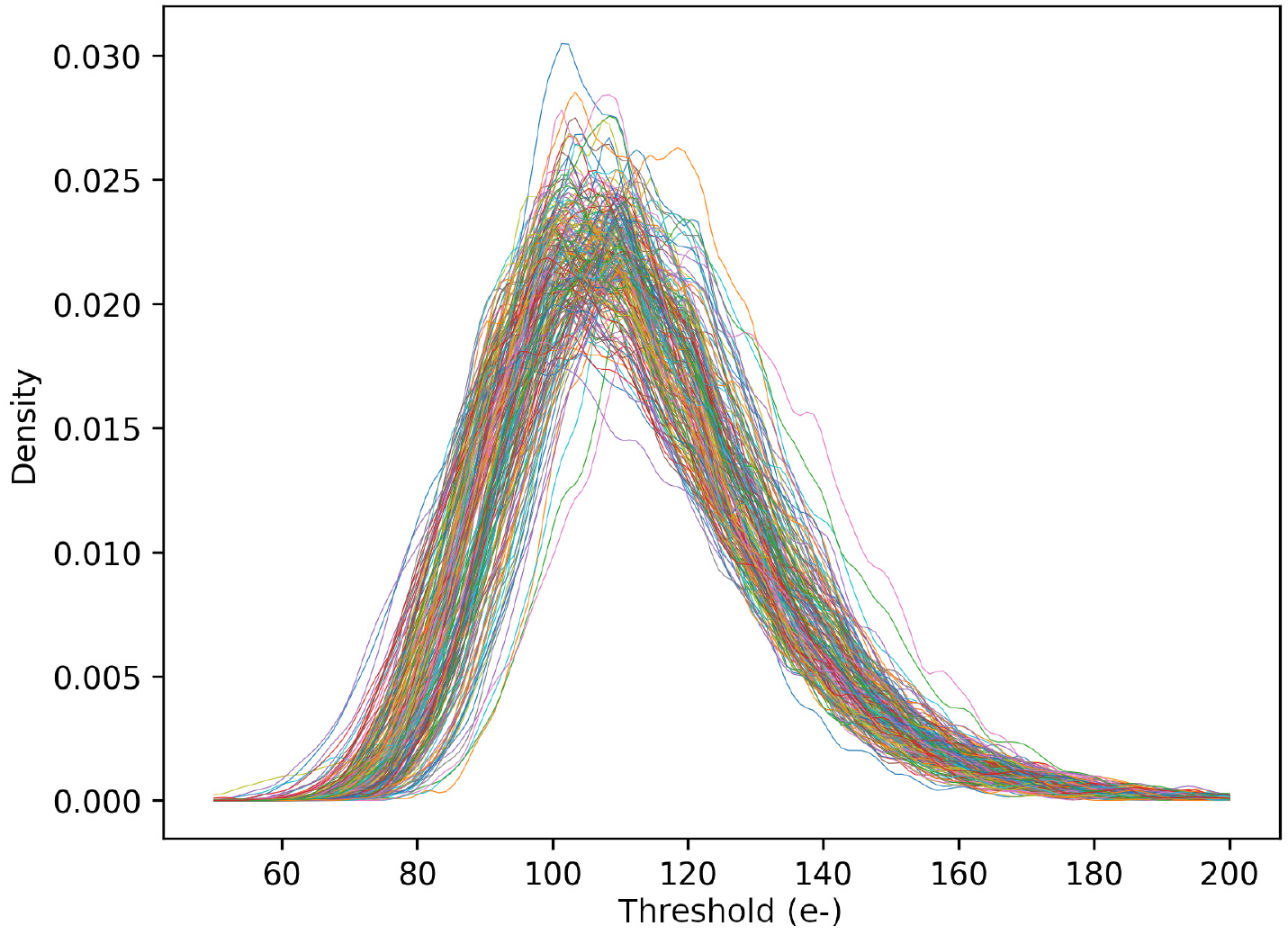}
  \caption{Probability density function of chip thresholds for each chip in one OL stave (the last stave shown in Fig.\,\ref{fig:tune}). Each line represents a chip. The average threshold is $113\,\mathrm{e^-}$ and the root mean square is $20\,\mathrm{e^{-}}$.}
  \label{fig:chips}
\end{figure}

Figure \ref{fig:chips} shows the threshold values of each chip in one stave. The average threshold across this stave is $113\,\mathrm{e^-}$ and the root mean square of thresholds per chip is $20\,\mathrm{e^{-}}$.\par

The full services, including the full data path described in Section \ref{section:readout}, as well as the final cooling system, were used to obtain these results.
\par

\FloatBarrier

\FloatBarrier

\subsection{Outer barrel fake hit rate}
\label{subsection:fhr}
\begin{figure}[htbp!]
  \centering
  \includegraphics[width=\textwidth]{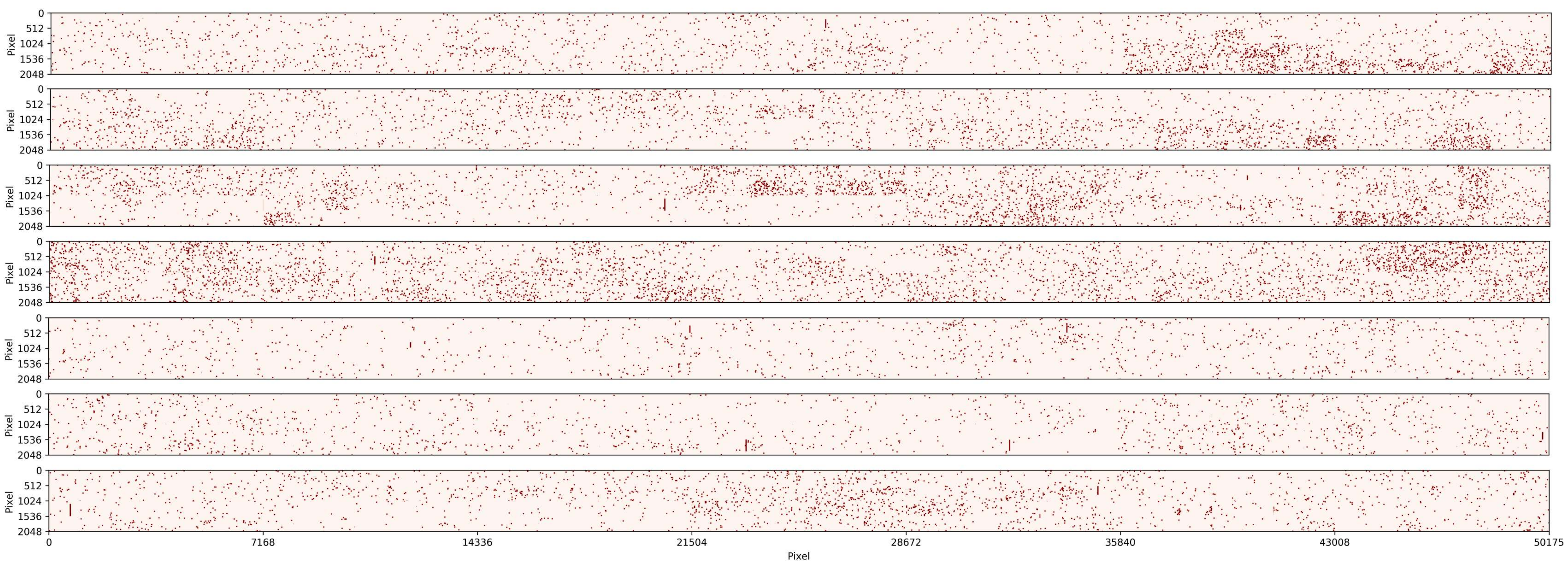}
  \caption{Fake hit scan of a selection of OL staves. Red markers denote hits. Each marker is enlarged by a factor of 40 for visibility.}
  \label{fig:fhr}
\end{figure}

The fake hit scan involves reading out hit pixels without any stimulation of the chip. A hit is due to either noise or a cosmic ray. The fake hit scan shown in Fig.\,\ref{fig:fhr} was performed after threshold tuning. The scan was performed over 5 minutes with $9.2 \times 10^5$ triggers. Without masking pixels, the fake hit rate of the staves shown was $3.8 \times 10^{-6}$\,/pixel/event, comparable to the requirement of $10^{-6}$\,/\,pixel\,/\,event. This leads to an occupancy of $0.5\,$/\,event\,/\,$\mathrm{cm^{2}}$, roughly the same as the particle hit density in the OB for central Pb-Pb collisions \cite{a}. The expected cosmic muon rate is $1\,\mathrm{cm^{-2}min^{-1}}$ \cite{f} which leads to a cosmic muon hit density of roughly $5.4 \times 10^{-6}\,$/\,event\,/\,$\mathrm{cm^{2}}$, $10^5$~times smaller than the fake hit occupancy. Some vertical lines of hits can be seen. These correspond to bad double columns, read out by a single broken priority encoder. 
\par

The number of hits varies across each chip within each stave. For the stave shown in Fig.\,\ref{fig:chips} for example, the total number of hits was roughly $25 \times 10^{6}$. These hits were from only 3700 pixels, that is a fraction of $10^{-5}$ of the total pixels on the stave. Of the hit pixels, 60\% had less than 3 hits and 80\% had less than 100 hits. See Table \ref{tab:hits} for an overview of the number of hits seen by each hit pixel. 19 pixels had a number of hits roughly equal to the number of triggers. These are known as stuck pixels, as they read out a hit on every trigger. The fake hit rate of this stave was $3 \times 10^{-7}$\,/\,pixel\,/\,event with all pixels included and $8 \times 10^{-8}$\,/\,pixel\,/\,event after the 19 stuck pixels were removed. The number of hit pixels per chip for this stave was on average 15 with a spread of 11. 

\begin{table}
  \centering
  \caption{Number of pixels with a given number of hits. Data from a single OL stave shown in Fig.\,\ref{fig:chips}.}
  \label{tab:hits}
  \begin{tabular}{ c  c  c }
    \hline
    Number of hits & Pixel firing probability & Number of pixels \\
    \hline
    0 & 0\% & 102756975 \\
    $1$ to $10^{2}$ & 0.01\% & 3106 \\
    $10^{2}$ to $10^{3}$ & 0.01\% to 0.1\% & 303 \\
    $10^{3}$ to $10^{4}$ & 0.1\% to 1\% & 197 \\
    $10^{4}$ to $10^{5}$ & 1\% to 10\% & 101 \\
    $10^{5}$ to $9.2 \times 10^{5}$ & 10\% to 99\% & 117 \\
    $\geq 9.2 \times 10^{5}$ & 100\% & 19 \\
    \hline
  \end{tabular}
\end{table}

\FloatBarrier

\section{Summary}
The replacement of the ITS during Long Shutdown 2 (LS2) will extend the physics reach of ALICE to lower transverse momentum, allowing the characterisation of the QGP via measurements of unprecedented precision. The use of a sole MAPS design, ALPIDE, is a huge step forward in terms of material budget, readout rate and spatial resolution. The new ITS is now fully constructed after a huge effort from numerous sites around the world. Notably, a yield of over 94\% was achieved for the OB staves. The readout chain for both the IB and OB has been demonstrated to work and a campaign to gather the first particle tracks from cosmic muons with the OB is ready to begin. The detector is on track for an installation in the ALICE cavern during LS2.


\begin{thebibliography}{99}

\bibitem{z}
  ALICE Collaboration, \emph{The ALICE experiment at the CERN LHC}, \emph{JINST 3 (2008) S08002}.

\bibitem{a}
  ALICE Collaboration, \emph{Technical Design Report  for the Upgrade of the Inner Tracking System}, \emph{J. Phys. G 41 (2014) 087002 [CERN-LHCC-2013-024]} (2013).

\bibitem{y}
  M. Mager on behalf of the ALICE Collaboration, \emph{ALPIDE, the Monolithic Active Pixel Sensor for the ALICE ITS upgrade}, \emph{Nucl.Instrum.Meth.A 824 (2016) 434-438}.

\bibitem{b}
  Tower Jazz, \emph{Specialty IC Manufacturing for a Smart World}, (2016), www.jazzsemi.com.
  
\bibitem{c}
  G. Aglieri Rinella on behalf of the ALICE Collaboration, \emph{The ALPIDE pixel sensor chip for the upgrade of the ALICE Inner Tracking
    System}, \emph{Nucl. Instrum. Meth. A 845 (2017) 583}.

\bibitem{x}
  J. Schambach et al. on behalf of the ALICE Collaboration, \emph{{{ALICE} inner tracking system readout electronics prototype testing with the {CERN} `Giga Bit Transceiver'}}, \emph{JINST 11 (2016) C12074}.
  
\bibitem{d}
  P. Moreira et al., \emph{The GBT project, in proceedings of Topical Workshop on Electronics for Particle
    Physics}, Paris, France, September 21-25 2009, pp. 342-346. 

\bibitem{e}
  A.Szczepankiewicz on behalf of the ALICE Collaboration, \emph{Readout of the upgraded ALICE-ITS}, \emph{Nucl. Instrum. Meth. A 824 (2016) 465-469}.

\bibitem{f}
  K. Nakamura et al. (Particle Data Group), \emph{J. Phys. G 37, 075021 (2010)}.
  


\end{thebibliography}
\end{document}